\documentclass[twocolumn,prl,amssymb,showpacs]{revtex4}
\usepackage{amsmath}

\def\vs{\medskip}

\newcommand{\ket}[1]{\mbox{$| #1 \rangle$}}
\newcommand{\bra}[1]{\mbox{$\langle #1 |$}}

\def\zu{\{0,1\}}

\def\r{_\mathrm{r}}
\def\s{_\mathrm{s}}
\def\c{_\mathrm{c}}

\def\a{\mathbf{a}}
\def\b{\mathbf{b}}
\def\x{\mathbf{x}}
\def\y{\mathbf{y}}
\def\k{\mathbf{k}}
\def\zero{\mathbf{0}}
\def\uns{\bra{1\mathrm{s}}}
\def\P{P_{\a, \b| \x ,\y}}
\def\chsh{\mbox{\small CHSH}}

\begin{document}

\title{Universally-composable privacy amplification from causality constraints}

\author{Llu\'{\i}s Masanes}
\affiliation{ICFO-Institut de Ciencies Fotoniques, Mediterranean Technology Parck, 08860 Castelldefels (Barcelona), Spain}

\date{\today}

\begin{abstract}
We consider schemes for secret key distribution which use as a resource correlations that violate Bell inequalities. We provide the first security proof for such schemes, according to the strongest notion of security, the so called universally-composable security. Our security proof does not rely on the validity of quantum mechanics, it solely relies on the impossibility of arbitrarily-fast signaling between separate physical systems. This allows for secret communication in situations where the participants distrust their quantum devices.
\end{abstract}
\maketitle

In an experimental set-up where a Bell inequality \cite{Bell} is violated, one has the certainty that the outcomes of some local measurements are not determined beforehand. This limits the amount of correlations between such outcomes and other systems not involved in the experiment. It also limits the knowledge about these outcomes that a distant party can have. This fundamental piece of our understanding of physical reality can be exploited for implementing information-theoretic tasks. For instance, in this letter we show that a secret key generated from the outcomes of Bell-violating measurements is secure. This reasoning is independent of quantum mechanics, the only key assumption is the impossibility of arbitrarily-fast signaling between separate systems.

\vs The first scheme for generating secret key from Bell-violating correlations was presented in \cite{BHK}, and was followed by others \cite{AGM, SGBMPA, AMP}. All these schemes where presented with partial security proofs. The results presented in this letter, complemented with the ones in \cite{us}, provide a general security proof without assumptions (apart from no signaling) for all these schemes. We use the strongest security criterion, the so-called universally-composable security \cite{C}, which warrants that key distribution is secure in any context. Our methods are very general, and can be adapted to other Bell inequality-based key-distribution schemes.


\vs{\em No signaling } Consider two parties, Alice and Bob, each having a physical system which can be measured with different observables. Let $a(b)$ be the outcome when Alice(Bob)'s system is measured with one of the observables parametrized by $x(y)$, with joint conditional probability distribution denoted by $P_{a,b|x,y}$. We say that $P_{a,b|x,y}$ is a nonsignaling distribution if the marginals depend only on their corresponding observables, that is $P_{a|x,y} = P_{a|x}$ and $P_{b|x,y} = P_{b|y}$ for all $a,b,x,y$ \cite{MAG}. It is clear that if one of these conditions is not satisfied, then arbitrarily-fast signaling is possible.

\vs{\em Nonlocality } The distributions that can be written as
\begin{equation}
  P_{a,b|x,y} = \mbox{$\sum_\lambda$} P_\lambda\, P_{a|x,\lambda}\, P_{b|y,\lambda}
\end{equation}
are called local, and satisfy all Bell inequalities \cite{MAG}. In the binary case ($a,b,x,y \in\zu$) all Bell inequalities are equivalent to the CHSH inequality \cite{CHSH}. For what follows, it is convenient to write the CHSH inequality as
$\bra{\chsh} P_{a,b|x,y} \rangle \geq \sqrt{2}$,
where the vector
\begin{equation}\label{chsh}
  \ket{\chsh} = \frac{1}{4\sqrt{2}}\
  \begin{array}{|cc|cc|}
      \hline
      1 & 5 & 1 & 5 \\
      5 & 1 & 5 & 1 \\
      \hline
      1 & 5 & 5 & 1 \\
      5 & 1 & 1 & 5 \\
      \hline
  \end{array}
\end{equation}
contains the coefficients of the inequality, and the vector
\begin{equation}
  \ket{P_{a,b|x,y}} =\
  \begin{array}{|cc|cc|}
      \hline
      P_{0,0|0,0} & P_{0,1|0,0} & P_{0,0|0,1} & P_{0,1|0,1} \\
      P_{1,0|0,0} & P_{1,1|0,0} & P_{1,0|0,1} & P_{1,1|0,1} \\
      \hline
      P_{0,0|1,0} & P_{0,1|1,0} & P_{0,0|1,1} & P_{0,1|1,1} \\
      P_{1,0|1,0} & P_{1,1|1,0} & P_{1,0|1,1} & P_{1,1|1,1} \\
      \hline
  \end{array}
\end{equation}
contains the probabilities for all experimental settings. [We arrange the components of these vectors in a table for the sake of clarity.] Notice that in this form, the lower the quantity $\bra{\chsh} P_{a,b|x,y} \rangle$ the larger the violation. The distribution attaining maximal violation ($\bra{\chsh} P_{a,b|x,y} \rangle =1/\sqrt{2}$) is the so-called PR-box \cite{PR}, which can be considered the maximally nonlocal (nonsignaling) distribution. The correlations generated by measuring quantum systems are constrained by Cirel'son's bound $\bra{\chsh} P_{a,b|x,y} \rangle \geq 2^{-1/2} 3 -1 \approx 1.121$ \cite{TB}.

\vs {\em Privacy amplification} (PA) is the procedure by which a partially secret $N\r$-bit string $\a$ (the {\em raw key}) is transformed into a highly-secret $N\s$-bit string $\k$ (the {\em secret key}) \cite{gpa}. Usually, the secret key is shorter than the raw key ($N\s < N\r$), which is the price for the gain in privacy. The function implementing this transformation $h(\a) = \k$ is called {\em hash function}. It is usually the case that the hash function has to be generated randomly after the raw key $\a$ has been obtained, but in our scheme, $h$ is fixed from the beginning and known to everybody, including the eavesdropper (Eve). An {\em ideal secret key} is a uniformly-distributed random variable $\k$ which is uncorrelated with the rest of the universe (Eve). The information held by Eve is encoded in the state of a physical system, which can be measured with one of many different observables, parametrized by $z$. If $P_{e|z}$ is the distribution for the outcomes when this system is measured with the observable $z$, then the distribution of an {\em ideal secret key} is $P^\mathrm{ideal}_{\k,e|z} = 2^{-N\s} P_{e|z}$. Usually, the {\em real secret key} generated by PA is not guaranteed to be an {\em ideal secret key},  $P_{\k,e|z} \neq 2^{-N\s} P_{e|z}$.

In general, PA constitutes a sub-routine within cryptographic protocols, which use secret key as an ingredient (an example being the encryption of messages).
It is desirable that the result obtained when any of these protocols is fed with the {\em real secret key}, is the same as if fed with an {\em ideal secret key}, with arbitrarily high probability. If this is the case, then we say that PA is universally composable, because it is secure in any context. Clearly, this happens if the {\em real} and {\em ideal} secret keys are indistinguishable.

The most general strategy for distinguishing the bipartite states $P_{\k,e|z}$ (the real key) and $2^{-N\s} P_{e|z}$ (the ideal key) consists of performing joint measurements on the key and Eve's system. The no signaling formalism alone does not say anything about joint measurements. However, the key is a classical system which can be observed without disturbing the global state. Therefore, the most general strategy is to read $\k$ and chose an observable $z$ depending on its value. It is well known that the probability of guessing correctly with the optimal strategy is
\begin{equation}\label{p succ}
  p_\mathrm{correct} = \frac{1}{2}+ \frac{1}{4} \sum_\k \max_{z} \sum_e \left| P_{\k, e|z} - 2^{-N\s} P_{e|z} \right| \ .
\end{equation}
Notice that the maximization on $z$ depends on $\k$. When (\ref{p succ}) is close to $1/2$, the optimal strategy for distinguishing the real from the ideal key is as good as a random guess---this is the security condition that we consider.

\vs In {\em key distribution} from Bell-violating correlations, Alice has $N$ systems, Bob has $N$ systems and, without loss of generality, Eve has one ``big" system, jointly distributed according to an arbitrary (unknown) $P_{\a, \b, e|\x ,\y ,z}$. [Bold symbols correspond to bit-string variables.] It is usually assumed that this is a $(2N+1)$-partite nonsignaling distribution \cite{us} ({\em i.e.} the marginals only depend on their corresponding observables), however, we are able to proceed with a weaker assumption. If the secret key is a function of Alice's string $\k= h(\a)$, then Bob's $N$ systems can be considered as a single ``big" system, that is, no-signaling within Bob's systems is not required in our proof. We refer to this assumption as ``$(N+2)$-partite no signaling". According to \cite{WR}, the even weaker assumption of 3-partite no signaling (where Alice's $N$ systems are also considered as single one) is insufficient to warrant security.
Of these $N$ pairs of systems, $N\r\ (N\r <N)$ are used for generating the raw key, and the rest are used to estimate how much nonlocality is shared by Alice and Bob \cite{us}. In the large-$N$ limit, $N\r$ is equal to $N$ up to terms sublinear in $N$---this is denoted by $N\r \approx N$.

The following result establishes the security of Alice's key $\k= h(\a)$ when $\a$ is generated by measuring $N\r$ of Alice's systems with the observable $x=0$. Of course, it is necessary that the correlations shared by Alice and Bob $\P$ have a sufficiently small value of $\bra{\chsh}^{\otimes N\r} \ket{\P}$, or in other words, are sufficiently nonlocal. However, the goal of key distribution is that both, Alice and Bob, hold the secret key $\k$. Later we address this problem.

\vs\noindent {\bf Main result } For almost all functions $h: \zu^{N\r} \rightarrow \zu^{N\s}$ and any $(N\r +2)$-partite nonsignaling distribution $P_{\a, \b, e|\x ,\y ,z}$, the random variable $\k = h(\a)$ satisfies
\begin{eqnarray}\nonumber
  && \sum_\k \max_{z} \sum_e \left| P_{\k, e|\x= \zero, z} - 2^{-N\s} P_{e|z} \right|
  \\ \label{result}
  &\leq& \sqrt{2}^{N\s+\sqrt{N\r}}\ \bra{\chsh}^{\otimes N\r} \ket{\P}\ ,
\end{eqnarray}
where $\zero$ is the zero vector.

\vs Here and in the rest of the letter we say that ``almost all functions have a particular property" if when randomly picking a function $h$ with uniform distribution over all functions $h: \zu^{N\r} \rightarrow \zu^{N\s}$ then the probability that $h$ does not have that particular property is lower than $2 \exp{\!(5 N\r-2^{\sqrt{N\r}}/4)}$. The above result is also true for any $\x \neq \zero$, but for simplicity we only consider the case $\x = \zero$, which is sufficient for key distribution.

When the given correlations $\P$ are generated by measuring quantum systems Cirel'son's bound implies $\bra{\chsh}^{\otimes N\r} \ket{\P} > 1$, which prevents the right-hand side of (\ref{result}) to be small. Hence, this simple scheme does not work with quantum correlations. This problem is solved by the BHK protocol, which yields large secure secret keys. The BHK protocol is analyzed below. Now, we proceed to prove the main result, and start by stating two lemmas which are proven in the Appendix.

\vs\noindent{\bf Lemma 1 } For any $(N\r+1)$-partite nonsignaling distribution $\P$  we have $P_{\a|\x=\zero} = \langle \Gamma_\a \ket{\P}$, where $\ket{\Gamma_\a} = \ket{\gamma_{a_1}} \otimes \cdots \otimes \ket{\gamma_{a_{N\r}}}$ and
\begin{equation}\nonumber
  \ket{\gamma_0} = \frac{1}{8}\
  \begin{array}{|rr|rr|}
      \hline
      1 &-3 & 1 & 5 \\
      5 & 1 &-3 & 1 \\
      \hline
      1 &-3 & 5 & 1 \\
      5 & 1 & 1 &-3 \\
      \hline
  \end{array}
  \ \ ,\ \
  \ket{\gamma_1} = \frac{1}{8}\
  \begin{array}{|rr|rr|}
      \hline
      1 & 5 & 1 &-3 \\
      -3& 1 & 5 & 1 \\
      \hline
      1 & 5 &-3 & 1 \\
      -3& 1 & 1 & 5 \\
      \hline
  \end{array}
  \ .
\end{equation}


\vs\noindent{\bf Lemma 2 } For any given function $h: \zu^{N\r} \rightarrow \zu^{N\s}$ and any $\k \in \{0,1\}^{N\s}$, define ${\cal A}_\k = h^{-1} (\k)$ and $\ket{ \Gamma_{\mathcal{A}_\k} } = \sum_{\a \in \mathcal{A}_\k} \ket{\Gamma_\a}$. Almost all functions $h$ satisfy
\begin{equation}\label{volem}
  \big| 2^{N\s}\, \ket{\Gamma_ {\mathcal{A}_\k}} - 4^{-N\r} \ket{1\mathrm{s}} \big| \preceq \sqrt{2}^{N\s+\sqrt{N\r}}\ \ket{\chsh}^{\otimes N\r}\ ,
\end{equation}
for all $\k$, where the symbol $|\cdot|$ denotes entry-wise absolute value, the symbol $\preceq$ denotes entry-wise less or equal than, and $\ket{1\mathrm{s}} \in \mathbb{R}^{16^{N\r}}$ has all entries equal to one.

\vs {\em Proof of the main result } Let $h$ be any of the functions which satisfies (\ref{volem}), and for each $\k$, let $\ket{\Gamma_{{\cal A}_\k}}$ be the vector defined in Lemma 2. Using $P_{\k|\x=\zero} = \langle\Gamma_{\mathcal{A}_\k} | \P\rangle$, the convexity of the absolute-value function, the inequality (\ref{volem}), and the fact that the marginal for $\a, \b$ cannot depend on $z$, we have
\begin{eqnarray}\nonumber
   && \sum_\k \max_{z} \sum_e P_{e|z} \Big| P_{\k |\x=\zero,e,z} - 2^{-N\s} \Big|
\\ \nonumber
  &\leq& \sum_\k \max_{z} \sum_e P_{e|z}\, \Big| \bra{\Gamma_{\mathcal{A}_\k}} - 2^{-N\s -2 N\r} \uns \Big| \ket{P_{\a, \b|\x ,\y ,e,z}}
\\ \label{ze}
  &=& \sqrt{2}^{N\s + \sqrt{N\r}} \bra{\chsh}^{\otimes N\r} \ket{\P}\ ,
\end{eqnarray}
which is precisely (\ref{result}).
\hfill $\Box$


\vs{\em Error correction and public communication }
It is usually the case that the given distribution $\P$ does not provide perfect correlations between $\a$ and $\b$. Hence, if $\a$ is the raw key, Bob has to correct the errors in $\b$ before applying the hash function $h$. This can be done by Alice publishing some information about $\a$, and Bob using it for correcting his errors. This is a standard procedure in quantum key distribution, which is detailed in \cite{us} or \cite{Renner}.
Other procedures within the key distribution protocol may also require public communication. Let the $N\c$-bit string ${\bf c}$ be all the information about $\a$ that Alice has published during the protocol. Because ${\bf c}$ is a function of $\a$, we can still use the main result (\ref{result}) in this new setting if we let both, $\k$ and ${\bf c}$, to be the outcomes of the function $h:\zu^{N\r} \rightarrow \zu^{N\c} \times \zu^{N\s}$. However, $\k$ and ${\bf c}$ play different roles: $\k$ is the secret key and ${\bf c}$ is part of the information owned by Eve. Hence, the extension of the security condition (\ref{result}) to the present setting is
\begin{eqnarray}\nonumber
  && \sum_{\k,{\bf c}} \max_{z} \sum_e \big| P_{\k,{\bf c},e| z} - 2^{-N\s} P_{{\bf c},e | z} \big|
  \\ \label{result comm}
  &\leq& 2\, \sqrt{2}^{N\c +N\s+ \sqrt{N\r}}\ \bra{\chsh}^{\otimes N\r} \ket{\P}\ ,\quad
\end{eqnarray}
where here and in the rest, the conditioning on $\x= \zero$ is implicit. This inequality is obtained by taking (\ref{result}) and using the triangular inequality with the third distribution $2^{-N\c -N\s} P_{e|z}$. The secret key is secure if the right-hand side of (\ref{result comm}) can be made arbitrarily small (as $N\r$ grows). This happens when the length of the final key is
\begin{equation}\label{rate}
  N\s \approx \log_{2}\!\Big[ \bra{\chsh}^{\otimes N\r} \ket{\P}^{-2}\Big] -N\c\ ,
\end{equation}
up to sub-linear terms.

\vs {\em Parameter estimation } In the unconditional-security scenario, the honest parties are given $N$ pairs of systems in a completely unknown global distribution. To perform a key distribution protocol, and in particular to set the numbers $N\s$ and $N\c$, they need to bound some quantities, like for instance $\bra{\chsh}^{\otimes N} \ket{\P}$. In order to do so, they invest some of the given pairs to obtain information about the distribution $\P$ of the $N\r$ remaining pairs. More precisely, they compute the bounds for $N\s, N\c$ for another distribution $P_{\a,\b,e|\x,\y,z}'$, which is warranted to be close to the real (unknown) one ($\sum_{\a,\b,e} | P_{\a,\b,e|\x,\y,z}' - P_{\a,\b,e|\x,\y,z} | \leq \epsilon$ for all $\x, \y$). This is explained with full detail in \cite{us}. It is shown in \cite{arxiv} that
\begin{eqnarray}\label{result smooth}
  && \sum_{\k, {\bf c}} \max_z \sum_e \big| P_{\k, {\bf c}, e|z} - 2^{-N\s} P_{e, {\bf c}|z} \big|
\\ \nonumber
  &\leq& 2\, \sqrt{2}^{N\s +N\c +\sqrt{N\r}}\ \bra{\chsh}^{\otimes N\r} \ket{\P'} + 2\epsilon\ ,
\end{eqnarray}
which provides the security bound for the real (unknown) distribution in terms of properties of any $\epsilon$-close primed distribution.


\vs{\em The BHK protocol} introduced in \cite{BHK} and analyzed in \cite{AMP,us} gives a rate of one secret bit per singlet ($\ket{00}+\ket{11}$). It is remarkable that this protocol, where the adversary is only constrained by no signaling, gives the same rate as if the adversary is constrained by no signaling plus quantum mechanics. The essential novelty of the BHK protocol is to measure each system with $m\geq 2$ observables, $x \in \{1,\ldots m\}$. In this case, instead of the CHSH, we use the Braunstein-Caves Bell inequality \cite{BC}, which can be expressed as $\langle \mbox{\small BC} \ket{P_{a,b|x,y}} \geq \sqrt{2}$, with
\begin{equation}\label{bc}
  \ket{\mbox{\small BC}} = \frac{1}{2\sqrt{2}\, m}\,
  \begin{array}{|cc|cc|cc|cc|cc|}
    \hline
    1 & \alpha & 1 & \alpha & & & & & & \\
    \alpha & 1 & \alpha & 1 & & & & & & \\
    \hline
    & & 1 & \alpha & 1 & \alpha & & & & \\
    & & \alpha & 1 & \alpha & 1 & & & & \\
    \hline
    & & & & \ddots & & \ddots & & & \\
    & & & & & & & & & \\
    \hline
    & & & & & & \ddots & & 1 & \alpha \\
    & & & & & & & & \alpha & 1 \\
    \hline
    \alpha & 1 & & & & & & & 1 & \alpha \\
    1 & \alpha & & & & & & & \alpha & 1 \\
    \hline
  \end{array}\ \ ,
\end{equation}
where $\alpha = 2 m +1$, and the empty entries represent zeroes. Notice that for $m=2$ this is equivalent to the CHSH inequality (\ref{chsh}). Following the same methods as above, one can prove inequalities analogous to (\ref{result}), (\ref{result comm}), (\ref{result smooth}), and obtain a key rate as in (\ref{rate}) but with the Braunstein-Caves Bell inequality
\begin{equation}\label{rate m}
  N\s \approx  \log_2\!\Big[ \bra{\mbox{\small BC}}^{\otimes N\r} \ket{\P}^{-2} \Big] -N\c\ .
\end{equation}
This rate formula can be improved by modifying $\ket{\mbox{\small BC}}$ in the following way: take the expression (\ref{bc}) and substitute $\alpha$ by $\sqrt{1+4m^2}$. The security of this rate will be proven somewhere else.

If Alice and Bob share singlets or something close to it, in the estimation process they measure them with all the observables corresponding to points in the equator of the block sphere (see \cite{BHK, AMP, us} for details), the generated correlations have $\bra{\mbox{\small BC}}^{\otimes N\r} \ket{\P} \approx 1/\sqrt{2}$, for large $m$. The raw keys $\a, \b$ are generated by measuring all systems with the same observable $x=0$, then $\a = \b$ and $N\c \approx 0$. Formula (\ref{rate m}) tells that the secret key rate is {\em one secret bit per singlet}: $N\s \approx N\r$. This rate cannot be improved because it is also the optimal rate achievable against a much weaker (quantum) adversary.

\vs{\em Conclusions } We show, for the first time, that key distribution from Bell-violating correlations is secure according to the strongest notion of security, the so called universally-composable security. This provides the possibility of implementing secure cryptographic protocols with untrusted quantum devices \cite{AGM, no trust}. In this model, Alice and Bob have to trust some of their apparatuses (classical computers and the random number generator), but can distrust the devices for preparing and measuring the quantum systems sent through the channel. The efficiency rate is slightly lower than the one obtained in standard quantum key distribution, where trusting the quantum devices is necessary.

Interestingly, in our scheme, Bell-inequality violation plays the same role as the min entropy \cite{Renner} does in standard quantum key distribution. Specifically, equations (\ref{result}) and (\ref{rate}) have a quantum counterpart, obtained with the exchange
\begin{equation}
  \log_2\! \Big[ \bra{\chsh}^{\otimes N} \ket{\P}^{-2} \Big] \ \ \leftrightarrow\ \ H_\mathrm{min} (\a|e)\ .
\end{equation}

A novelty of our scheme is that randomness extraction, or equivalently PA, can be performed with a constant hash function. This contrasts with previous methods for extracting randomness (two-universal hashing \cite{gpa}, extractors, etc.), which need random functions. However, we still lack an explicit construction for one of such hash functions.

\vs{\em Acknowledgements } The author is grateful to Renato Renner and Andreas Winter for valuable comments. This work is supported by Caixa Manresa, the spanish MEC (FIS2005-04627, FIS2007-60182, Consolider QOIT), and the EU-IP programme SCALA.

\section{Appendix}

Here we show the two lemmas stated above.

\vs \noindent {\em Proof of Lemma 1.} Here we use the same tools as in the proof of Lemma 16 from \cite{us}. By definition we can write $P_{\a|\x = \zero} = \langle \Gamma'_\a \ket{\P}$, where $\ket{\Gamma'_\a} =$ \mbox{$\ket{\gamma'_{a_1}} \otimes \cdots \otimes \ket{\gamma'_{a_{N\r}}}$} and
\begin{equation}\nonumber
  \ket{\gamma'_0} =
  \left[
    \begin{array}{rr|rr}
      1 & 1 & 0 & 0 \\
      0 & 0 & 0 & 0 \\
      \hline
      0 & 0 & 0 & 0 \\
      0 & 0 & 0 & 0
    \end{array}
  \right]\ ,
\quad
  \ket{\gamma'_1} =
  \left[
    \begin{array}{rr|rr}
      0 & 0 & 0 & 0 \\
      1 & 1 & 0 & 0 \\
      \hline
      0 & 0 & 0 & 0 \\
      0 & 0 & 0 & 0
   \end{array}
  \right]\ .
\end{equation}
The fact that Bob (when considered as a single system) cannot signal to Alice can be expressed as $P_{\a|\x, \y} = P_{\a|\x, \y'}$ for any $\y,\y'$. This implies that $P_{\a|\x = \zero} = \langle \Gamma''_\a \ket{\P}$ where $\ket{\Gamma''_\a} = \ket{\gamma''_{a_1}} \otimes \cdots \otimes \ket{\gamma''_{a_{N\r}}}$ and
\begin{eqnarray*}
  \ket{\gamma''_0} = \frac{1}{4}
  \left[
    \begin{array}{rr|rr}
      2 & 2 & 2 & 2 \\
      0 & 0 & 0 & 0 \\
      \hline
      -1 & -1 & 1 & 1 \\
      1 & 1 & -1 & -1
    \end{array}
  \right]\ ,
  \\
  \ket{\gamma''_1} = \frac{1}{4}
  \left[
    \begin{array}{rr|rr}
      0 & 0 & 0 & 0 \\
      2 & 2 & 2 & 2 \\
      \hline
      1 & 1 &-1 &-1 \\
      -1 & -1 & 1 & 1
   \end{array}
  \right]\ .
\end{eqnarray*}
The fact that each of the $N\r$ Alice's systems cannot signal to the rest, together with Bob's systems, implies the statement of the lemma. \hfill $\Box$

\vs {\em Proof of Lemma 2 } Within this proof, the entries of any vector $\ket{\Phi} \in \mathbb{R}^{16^{N\r}}$ are labeled as $\Phi (\a,\b,\x,\y)$. Also, for any pair of bit-strings $\x,\y$: (i) the string $\x\cdot\y$ is the bit-wise product, (ii) the string $\x\oplus\y$ is the bit-wise xor, and (iii) the integer $\|\x\|$ is the number of ones in $\x$. Using this notation we can write the entries of the vector $\ket{\chsh} ^{\otimes N\r}$ as $\chsh^{\otimes N\r} (\a,\b,\x,\y) = 2^{-5 N\r/2} 5^{\|\a\oplus\b\oplus \x\cdot \y\|}$. Next we prove inequality (\ref{volem}) for a given $\k$ and a given entry $(\a_0, \b_0 ,\x_0 ,\y_0 )$. Let $V_\a =1$ if the string $\a$ belongs to ${\cal A}_\k$, and $V_\a =0$ otherwise. If we pick a random function $h$ with uniform distribution over the set of all functions, then the random variables $V_\a$ are independent and distributed according to $\mbox{Prob}\{ V_\a = 1\} = 2^{-N\s}$, for all $\a$. Let $\mu_\a = \Gamma_\a (\a_0, \b_0 ,\x_0 ,\y_0 )$, $M=\|\a_0 \oplus\b_0 \oplus\x_0 \cdot\y_0 \|$, and note that $|\mu_\a| \leq 5^M 8^{-N\r}$ for all $\a$. Following Bernstein's contruction, for any $J$ and $\beta \geq 0$ we have
\begin{eqnarray}
  \nonumber
  && \mathrm{Prob} \left\{\sum_\a \mu_\a V_\a \geq J \right\}
  \\ \nonumber &\leq&
  \mathrm{Prob} \left\{ e^{-\beta J +\beta \sum_\a \mu_\a V_\a} \geq 1 \right\}	
  \\ \nonumber &\leq&
  e^{-\beta J} \left\langle e^{\beta \sum_\a \mu_\a V_\a} \right\rangle	
  \\ \nonumber &\leq&
  e^{-\beta J} \prod_\a \left[ 2^{-N\s} e^{\beta \mu_\a} + \left( 1-2^{-N\s} \right) \right]	
  \\ \label{expansion} &\leq&
  e^{-\beta J} \prod_\a \left[ 1+ 2^{-N\s} \left( \beta \mu_\a + \beta^2 \mu_\a^2 \right) \right]	
  \\ \nonumber &\leq&
  \exp{\!\big[-\beta J 2^{-N\s} \mbox{$\sum_\a$} (\beta \mu_\a + \beta^2 \mu_\a^2) \big]}
\end{eqnarray}
where in (\ref{expansion}) we need $|\beta\, 5^M 8^{-N\r}| \leq 1$. In this step we have used the expansion $e^x \leq 1+x+x^2$, which holds if $x \leq 1$. With a little work one obtains $\sum_\a \mu_\a = 4^{-N\r}$ and $\sum_\a \mu_\a ^2 \leq 2^{-5 N\r} 5^{2 M}$. Substituting this two expressions, $J= 2^{-N\s -2 N\r} + 2^{(\sqrt{N\r}- N\r-N\s)/2}\, 4^{-N\r}\, 5^M$ and $\beta = 2^{(\sqrt{N\r}+ N\r+N\s)/2}\, 4^{N\r}\, 5^{-M}$ we get
\begin{eqnarray*}
  \mathrm{Prob} \big\{ \mbox{$\sum_\a$} \mu_\a V_\a \geq 2^{-N\s -2 N\r} + 2^{(\sqrt{N\r}- N\r-N\s)/2}\, 4^{-N\r}\, 5^M \big\}
  \\
  \leq\ e^{-2^{\sqrt{N\r}}/4}\ .\ \
\end{eqnarray*}
Note that the chosen value for $\beta$ satisfies the required constraint. The expression obtained when replacing ``$\geq$" with ``$\leq$" above, can be derived in a similar way. Then
\begin{eqnarray}\nonumber
  && \big| 2^{N\s}\, \Gamma_ {\mathcal{A}_\k} (\a_0 ,\b_0 ,\x_0 ,\y_0 ) - 4^{-N\r} \big|
  \\ \label{vo}
  &>& \sqrt{2}^{N\s+ \sqrt{N\r}}\ \chsh^{\otimes N\r} (\a_0 ,\b_0 ,\x_0 ,\y_0 )
\end{eqnarray}
holds with probability $2\, e^{-2^{\sqrt{N\r}}/4}$.
However, we want this to not hold for all $\k$ and all entries $(\a,\b,\x,\y)$. The number of different values of $\k$ is $2^{N\s}$, and the number of different entries is $16^{N\r}$, then the probability for (\ref{volem}) being not true is upper-bounded by $2 \exp{\!(5 N\r -2^{\sqrt{N\r}}/4 )}$. \hfill $\Box$

\end{document}